\documentclass[11pt]{article}
\usepackage{newpasp}
\usepackage{graphicx}
\usepackage{epsf}

\def\edcomment#1{\iffalse\marginpar{\raggedright\sl#1\/}\else\relax\fi}
\reversemarginpar

\begin{document}
\title{(Sub-)mm interferometry in massive star-forming regions}
 \author{Henrik Beuther}
\affil{Harvard-Smithsonian Center for Astrophysics, 60 Garden Street, Cambridge, MA 02138, USA}

\begin{abstract}
(Sub-)mm interferometry is the most favorable technique to investigate
the earliest stages of massive star formation. I will outline general
applications in that field and discuss results of different sub-topics
(hot core chemistry and massive molecular outflows). Furthermore,
recent data obtained with the Submillimeter Array will be shown to
present the unique capabilities of this new instrument. Finally, I
will give a short outlook on the main physical topics of massive star
formation to be tackled with (sub-)mm interferometry within the next
decade.
\end{abstract}

\section{Introduction}

As high-mass star formation occurs only in a clustered mode and such
regions are on the average at a distance of a few kpc from the sun,
high spatial resolution is essential to investigate massive evolving
clusters. While more evolved clusters are obvious targets of optical
and infrared astronomy, this is different for the earliest stages of
their evolution. During their birth-phase high-mass star-forming
regions are deeply embedded within their natal cores with column
densities as high as $10^{25}$\,cm$^{-2}$, corresponding to visual
extinctions $A_{\rm{v}}\sim 10000$. Thus, they are not observable in
the near-infrared and even difficult in the mid-infrared (see
contribution by J. de Buizer, this volume.). Furthermore, hard X-ray emission
as observable with CHANDRA and XMM is only capable of penetrating an
$A_{\rm{v}}$ of around 100. The other {\it classical} approach to
investigate massive star-forming regions is studying their free-free
emission in the cm regime. But again, at their earliest stages massive
star-forming cores either do not emit free-free emission yet, or they
are confined by their surrounding cores that only weak or no
free-free emission can escape the region.

However, very young high-mass star-forming regions are strong
emitters of cold dust emission which is most prominent in the mm and
sub-mm wavelength regime. In addition, those regions exhibit a rich
molecular chemistry which again is best observed in the (sub-)mm
regime. Thus, combining the need for high spatial resolution and the
strength of young high-mass star-forming regions in the (sub-)mm
regime, (sub-)mm interferometry is the best tool to investigate the
physical processes in their earliest evolutionary phases. This short
review is meant to present the unique capabilities for studying
massive star formation with current and future (sub-)mm
interferometers. 

\subsection{General applications}

Because (sub-)mm interferometers are sensitive to continuum emission due to
the cold dust cores as well as to many molecular line transitions in the
(sub-)mm regime, possible fields of research are manyfold. As a very
short summary one could categorize some sub-applications:
\begin{itemize}

\item
Studies of the dust continuum emission to derive masses, column
densities, morphologies, multiplicities and eventually a proto-cluster
mass function.

\item
Investigations of the gas dynamics via studying the velocity structure of
different molecular line transitions.

\item
Studying massive proto-stellar disks via continuum and line studies.

\item
Investigations of the chemical evolution via molecular line studies.

\item
Studies of massive molecular outflows to get indirect evidence on the
physical processes taking place at the core centers.

\item
Identifications of unambiguous infall signatures via the line-profiles.

\end{itemize}

Those are not the only applications but these topics cover a
wide range of interesting questions.

\section{The hot core W3(H$_2$O)}

To give an example of gas and dust studies at high spatial resolution,
the case of W3(OH) is presented. VLA cm studies reveal that $6''$
offset from the main UCH{\sc ii} region W3(OH) another weaker cm
source associated with H$_2$O maser emission is found, dubbed
W3(H$_2$O) (Reid et al. 1995, Wilner et al. 1999). Based on the
spectral index in the cm regime, Reid et al. (1995) infer that the
emission is likely due to a synchrotron jet. Additional VLBI H$_2$O
maser observations show an outflow around W3(H$_2$O) (Alcolea et
al. 1992). Wyrowski et al. (1997,1999) observed this source
intensely in the mm regime with the Plateau de Bure Interferometer
(PdBI), and their results are quite astonishing. First of all, they
detect dust emission associated with the synchrotron jet. Furthermore,
they can disentangle two more dust continuum peaks less than 2
arc-seconds apart (Fig. 1a). This indicates that in the close
neighborhood of W3(OH) another massive cluster is evolving.

\begin{figure}[htb]
\includegraphics[width=7cm,angle=-90]{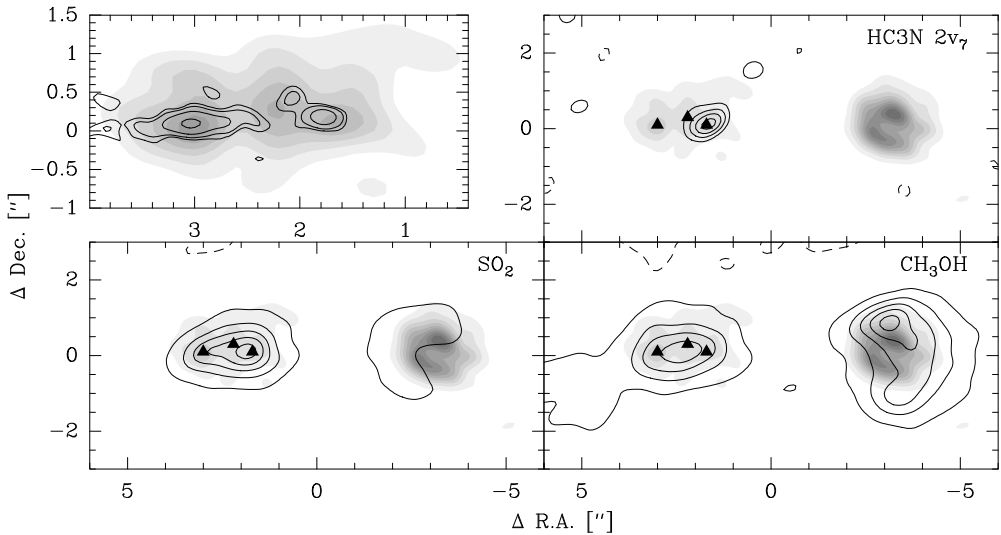}
\caption{\footnotesize PdBI mm images of W3(OH) by Wyrowki et
al. (1999). The top left panel shows a zoom into W3(H$_2$O) with the
mm dust continuum in grey-scale and the cm synchrotron jet in contours
(Wilner et al. 1999). The other three panels show the dust continuum
emission of the whole region W3(OH) in grey-scale and three different
molecular transitions as contour overlays. The molecules are marked at
the top right of each panel.}
\end{figure}

In addition to the continuum emission, they observed 6 different
molecular species simultaneously, Figure 1 shows three
examples of their images. It is striking that nitrogen-bearing species
like HC$_3$N are only observed toward W3(H$_2$O)~-- and there only
toward the western clump~-- whereas oxygen-bearing species like
CH$_3$OH or SO$_2$ are observed toward the young hot core source
W3(H$_2$O) as well as toward the more evolved UCH{\sc ii} region
W3(OH). Furthermore, based on HNCO observations Wyrowski et al. (1999)
estimate gas temperatures toward the HC$_3$N source around 200\,K,
clearly indicative of a hot core with an internal heating source. The
differences in oxygen- and nitrogen-bearing molecules are attributed
to evolutionary and chemical sequences. Better knowledge of the
chemical details will allow using different molecular species as
chemical clocks for star-forming regions. Wyrowski et al. (1999)
note an additional interesting feature, namely that nearly all
oxygen-bearing species peak north-west of W3(OH) whereas SO$_2$
peaks slightly east of it (Fig. 1).

\section{Massive molecular outflows}

\subsection{Scientific background}

One of the main questions in massive star formation over the last
decade is whether massive stars form in a similar manner as low-mass
stars, i.e., via disk accretion processes, or whether different
physical processes take place at the very dense evolving core centers,
namely the coalescence and merging of intermediate-mass proto-stars
(e.g., Wolfire et al. 1987, Bonnell et al. 1998, Stahler et
al. 2000, McKee \& Tan 2002). It is still difficult to disentangle 
high-mass core centers spatially well enough to infer this question in
a direct way. 

One very promising indirect approach is the observation of massive
molecular outflows. The accretion scenario needs massive proto-stellar
disks and thus predicts collimated molecular outflows. Contrary,
collisions of intermediate-mass proto-stars are presumably
extremely energetic and explosive, and it is unlikely that collimated
and well ordered structures can survive such processes.  Thus,
observing molecular outflows, and specifically their degree of
collimation, gives a handle on the physical processes taking place at
the cluster centers.

Several groups studied massive molecular outflows with single-dish
telescopes over recent years (e.g., Shepherd \& Churchwell 1996, Ridge
and Moore 2001, Zhang et al. 2001, Beuther et al. 2002a). All those
studies agree that massive outflows are ubiquitous phenomena in
massive star formation, and that they are far more massive and
energetic than their low-mass counterparts (see also Richer et
al. 2000). However, those studies disagree on the aspect of
collimation: Shepherd \& Churchwell (1996) and Ridge \& Moore (2001)
and observe on average lower degrees of collimation than known for
low-mass outflows. The coalescence community interprets these data as
indicative of different entrainment mechanisms (Stahler et al. 2000,
Churchwell 2000).

However, Beuther et al. (2002a) observed 26 massive molecular
outflows with the IRAM 30\,m telescope in the CO(2--1) transition with
a spatial resolution of $11''$. They observe a larger percentage of
bipolarity and collimation than previous studies, and they attribute
it to their better spatial resolution. To summarize their results:
their data are well consistent with outflows as collimated as their
low-mass counterparts, and no other physical processes need to be
invoked.

Nevertheless, even the spatial resolution of the latter observations
is too course too draw final conclusions, and the ultimate proof
whether massive collimated outflows do exist can only stem from
high-spatial-resolution interferometric observations.

\subsection{High-resolution observations: The case of IRAS\,05358+3543}

Different groups have embarked on observing massive molecular outflows
at high spatial resolution (e.g., Shepherd et al. 1998, Beuther et
al. 2002b, Gibb et al. 2003). However, I will focus mainly on the case of
IRAS\,05358+3543 observed with the Plateau de Bure Interferometer PdBI
(Beuther et al. 2002b). Previous single-dish observations reveal
strong wing emission, but determining the outflow morphology is difficult
based on the low-resolution data (Beuther et al. 2002a). The picture
changes considerably when the source is observed with the high spatial
resolution of the PdBI. We observed IRAS\,05358+3543 in two
setups, one focused on CO(1--0) and one observing the SiO(2--1) and
H$^{13}$CO$^+$(1--0) lines simultaneously. Figure 2 presents images of
all three lines as overlays on the infrared H$_2$ emission.

\begin{figure}
\includegraphics[width=8.3cm,angle=-90]{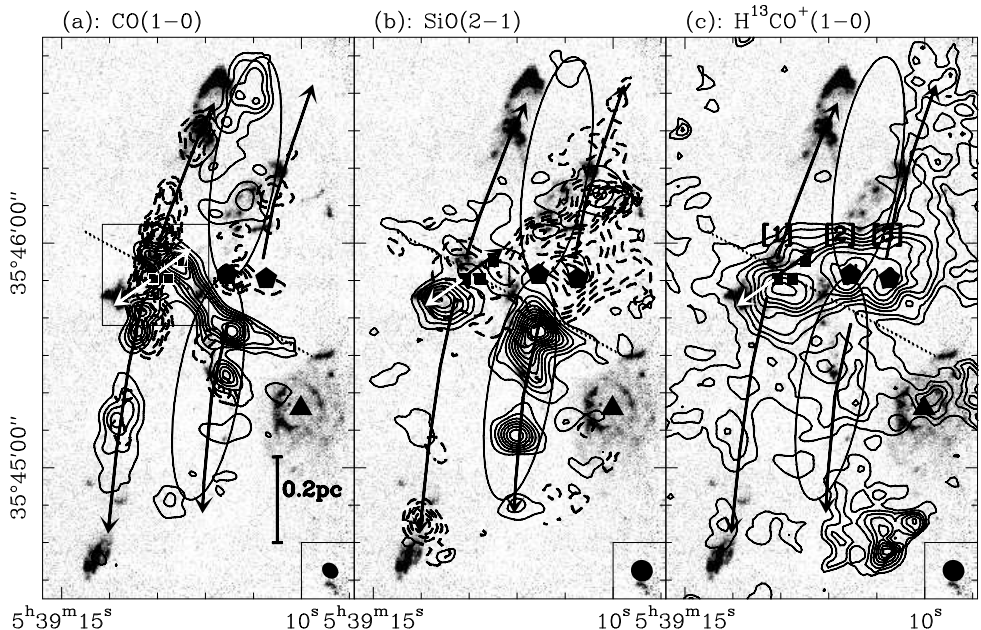}
\caption{\footnotesize Presented are the PdBI observations as contour 
overlays on the grey-scale H$_2$ data. The solid and dashed lines in
Figures {\bf (a) \& (b)} show the blue and red wing emission of the
CO(1--0) and SiO(2--1) emission, respectively. The solid lines in
Figure {\bf (c)} present the integrated H$^{13}$CO$^+$(1--0)
emission. The numbers in brackets label the three H$^{13}$CO$^+$
sources and the beams are show at the bottom left. In all images the
arrows and ellipses sketch the three outflows (the arrows to the right
and the ellipses represent two slightly different interpretations of
the western outflow). The three squares represent the three mm
sources, the diamonds locate the H$^{13}$CO$^+$ peaks, and the
triangle marks the IRAS 12\,$\mu$m position.}
\end{figure}

The main feature in Figure 2a is the highly collimated CO outflow in
the east with a collimation degree of 10, as high as the highest
observed in low-mass sources. It emanates from one of three
sub-condensations and terminates in the H$_2$ bow shocks in the north
and south. The second outflow further west~-- also in north-south
direction~-- is better depicted by the SiO data in Figure 2b. Due to the
mass-sensitivity limit around 50\,M$_{\odot}$ of the dust continuum
emission, no driving source is detected in the mm
continuum. Nevertheless, the H$^{13}$CO$^+$ data show a ridge of three
sources in east-west direction. One of those clumps is likely to
harbor the driving source of the second outflow. Furthermore, we
detect a third outflow inclined by 45$^{\circ}$ to the first flow and
emanating from the same mm-sub-clump (sketched by the two small
arrows), likely to harbor a double source. For a more detailed
description of the outflow morphologies and dynamics of IRAS
\,05358+3543 see Beuther et al. (2002b).

The important result one has to take away from those observations is
that highly collimated outflows do exist in massive star-forming
regions. The large-scale eastern outflow is difficult to explain in
the coalescence model, whereas it is consistent with a disk-accretion
scenario. Furthermore, complicated and seemingly strange features one
might observe with single-dish instruments can be disentangled with
high enough spatial resolution into features well known from low-mass
star formation. The source IRAS\,05358+3543 is complicated due to its
clustered mode of formation and interactions between members of the
evolving cluster, but regarding the physical processes we do not
observe outstanding differences from low-mass sources. Thus, these data
are well consistent with massive stars forming via {\it traditional}
disk-accretion processes. Beuther et al. (2002b) do not claim that
coalescence does not occur, they just infer that it does not seem to
be necessary.

The statistical database of high-spatial-resolution outflow
observations is still poor but slowly more data emerge which back the
case of IRAS\,05358+3543 (e.g., G35.2, Gibb et al. 2003;
IRAS\,20293+3952, Beuther et al., in prep.).

\section{Recent results from the Submillimeter Array}

As the Submillimeter Array (SMA) is getting online now, I like to
present its first results obtained in the field of massive star
formation.  The SMA is the first imaging interferometer in the sub-mm
wavelength regime, and its capabilities regarding high-mass star
formation are unique (Moran 1998).  Dust emission scales approximately
with $\nu^4$, thus the core emission increases quickly when observing
at shorter wavelength. Furthermore, massive hot cores are known to
exhibit molecular line forests (e.g., Schilke et al. 2000), and the
broad SMA correlator bandwidth of 2\,GHz in a double sideband mode
allows to image a multitude of molecular lines simultaneously. Thus,
it is possible to investigate outflows, disks, dust emission, gas
dynamics, and chemical processes throughout the same observations. For
more details about the SMA and its recent results see the
contributions by Zhang, Ohashi and Wilner in this volume.

In the following, I present recent SMA data observed toward the
massive star-forming region IRAS\,18089-1732 (Beuther et al., in
prep.). This region is in a very early evolutionary stage at the verge
of forming a hot core. Figure 3 shows spectra obtained at 217\,GHz and
344\,GHz with the current bandwidth of 1\,GHz (the full 2\,GHz will be
available within months). Obviously, we observe a forest of molecular
lines, and combining the 1\,mm and 850\,$\mu$m data of both sidebands
we identify 28 lines from 16 different molecular species.

\begin{figure}[htb]
\includegraphics[width=6.6cm]{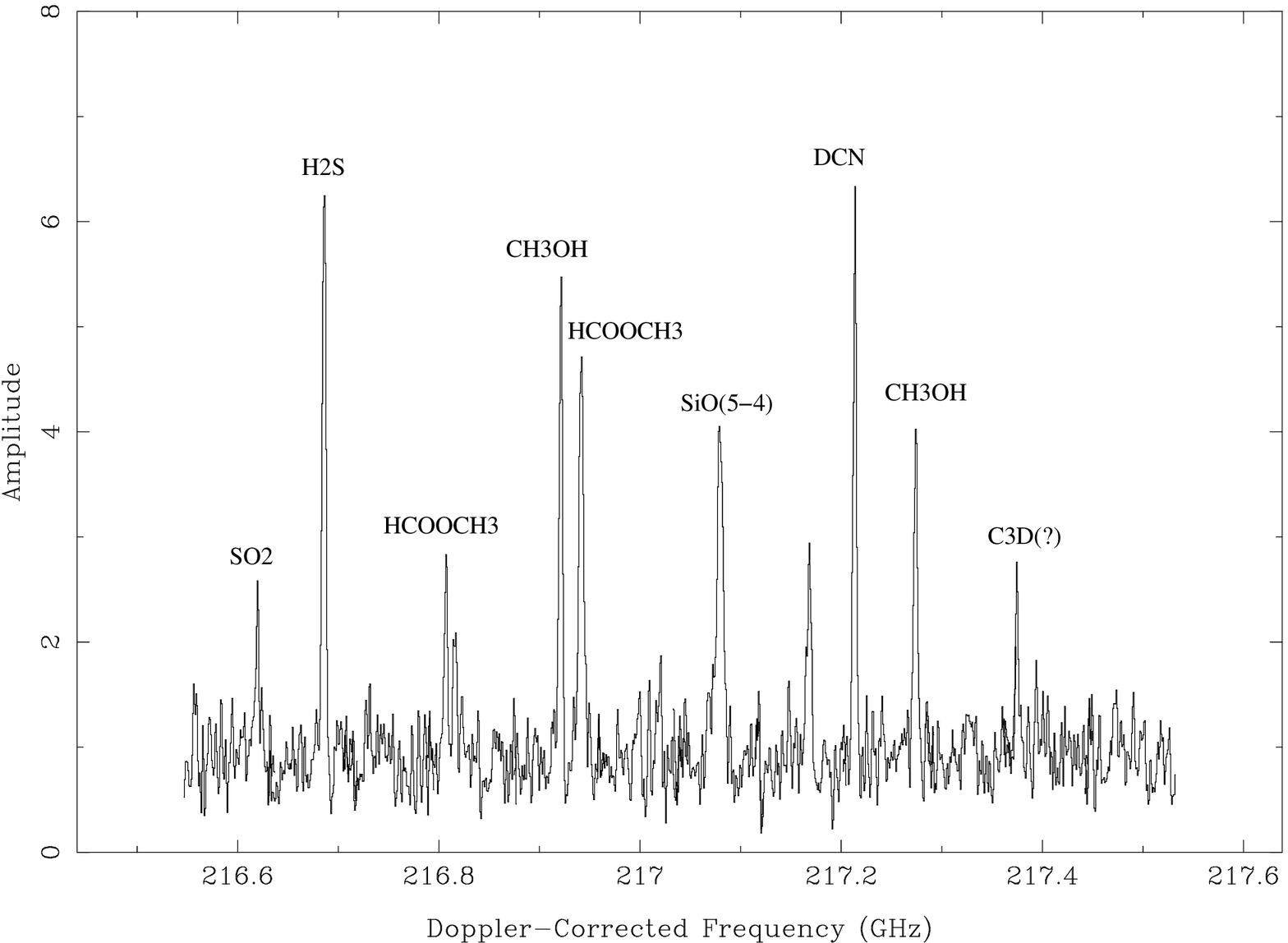}
\includegraphics[width=6.6cm]{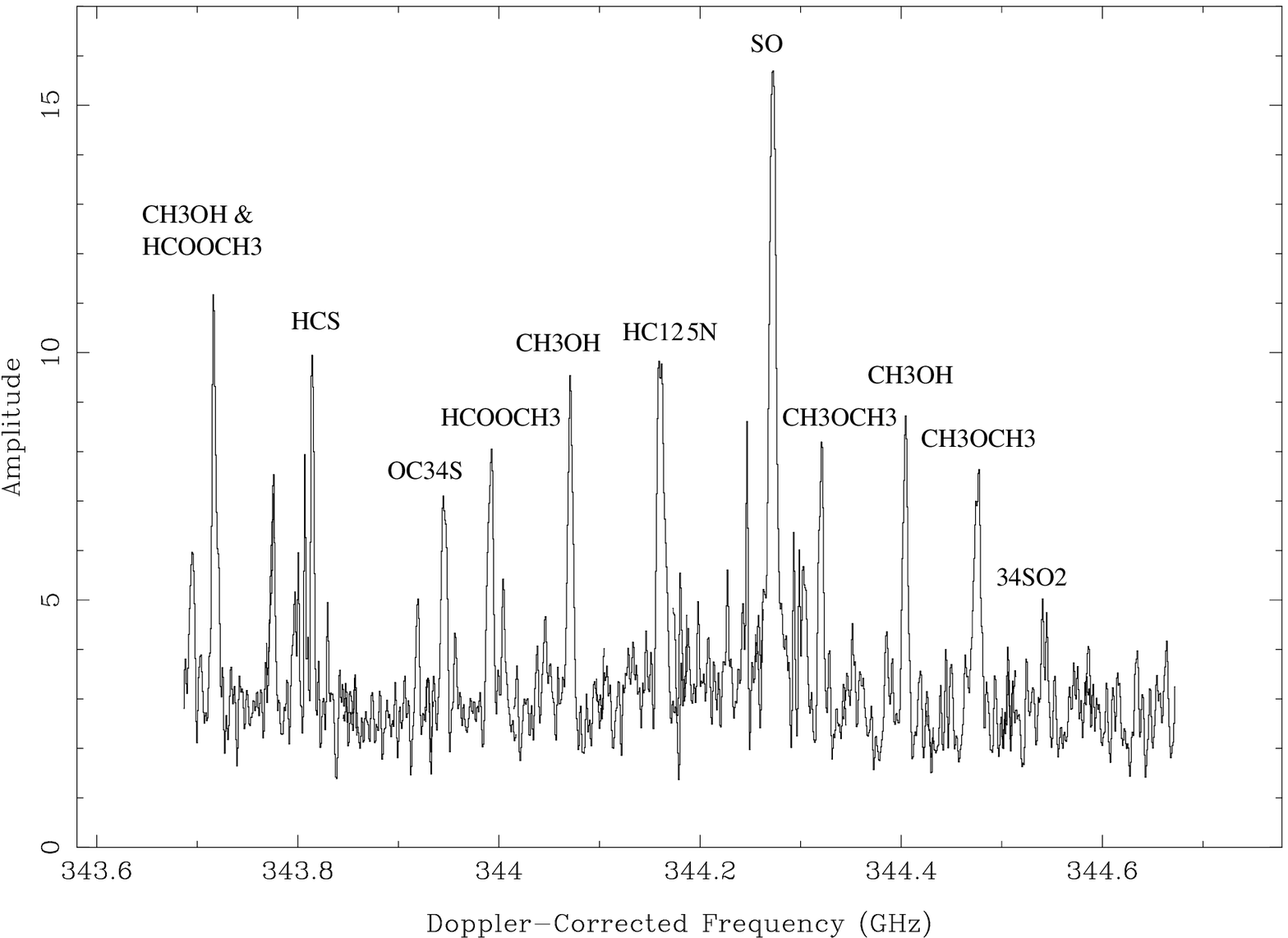}
\caption{\footnotesize SMA spectra of the high-mass star-forming
region IRAS\,18089-1732. The left spectrum is taken at 217\,GHz and
the right one at 344\,GHz.}
\end{figure}

Spectra are already interesting, but the excitement of interferometry starts
because of its capability to image all those lines simultaneously. 
Figures 4a and 4b show 2 representative examples of the imaged
molecular lines.

The 217\,GHz observations were centered at the SiO(5--4) line to
observe jet-like features from this region. Clearly, we identify a
collimated outflow emanating in the northern direction from the massive
core (Figure 4 left). The dust emission does not split up in multiple
sources, but it is resolved and we observe a core-halo
structure. Based on the dust emission, the core mass is estimated 
to be around 1400\,M$_{\odot}$.

\begin{figure}[htb]
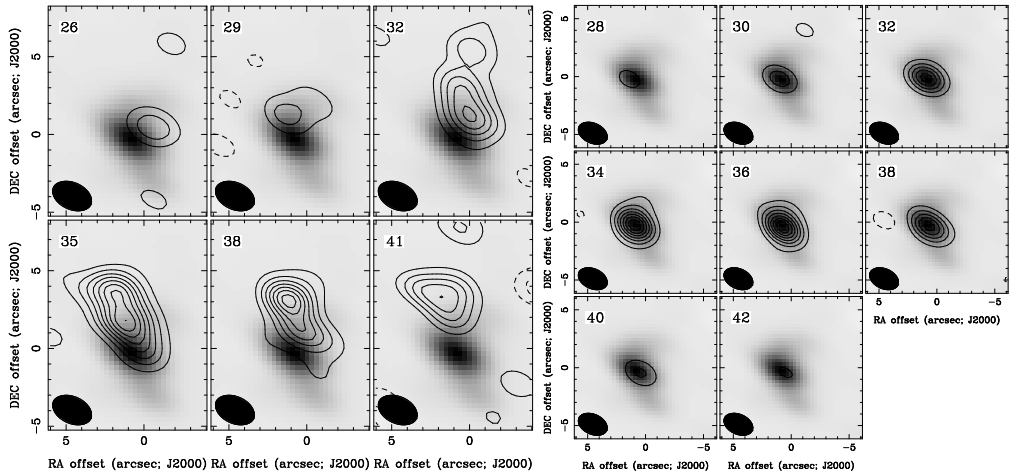

\includegraphics[width=6.2cm,angle=-90]{beuther_h_fig4a.ps}
\includegraphics[width=6.2cm,angle=-90]{beuther_h_fig4b.ps}
\caption{\footnotesize Contour channel maps for IRAS\,18089-1732 in
the SiO(5--4) (left) and HCOOCH$_3$ lines (right). The grey-scale
shows the 1.3\,mm continuum in all panels.}
\end{figure}

Figure 4 (right) shows the emission from the high density tracer HCOOCH$_3$
which~-- as expected~-- follows the core emission. Interestingly, we
find a velocity gradient in east-west direction roughly perpendicular
to the SiO outflow. The spatial scales of that velocity shift are
around 3500\,AU, and we interpret this velocity shift due to the
rotation of the outer parts of a disk. A similar velocity
pattern is observed in the CH$_3$OH line. Disk observations at even
higher spatial resolution are crucial to get a more consistent picture
of the massive star-forming processes.

Other molecules like H$_2$S, SO$_2$ or DCN show emission toward the
core as well as toward the outflow. It is likely that those molecules
are released and subsequently processed during shock interactions of
the outflow with the surrounding gas.

Already those preliminary results outline the unique capabilities of
the SMA, and in the years coming massive star formation research will
benefit tremendously from its various applications.

\section{The future of (sub-)mm interferometry}

Millimeter interferometry has proven to be a very powerful tool for
investigating the earliest stages of massive star formation. And the
future is even brighter: In addition to currently available
instruments like the PdBI, NMA and ATCA, the SMA is just at the
beginning of scientific research, and last but not least, within the
next decade two new instruments will be available, CARMA and
ALMA. Especially ALMA with its unprecedented sensitivity, spatial
resolution and imaging capabilities will revolutionize not just
massive star-forming studies but astrophysical research as a whole.

To conclude, I like to present some scientific topics in high-mass star
formation research which will be tackled and hopefully answered within
the next decade:

\begin{itemize}

\item
Detailed studies of massive disks.

\item
Chemistry of high-mass cores.

\item
Continuum and IMF studies of massive star-forming regions.

\item
Absorption and infall studies against the strong dust continuum at
very high frequencies.

\item
Studies of the warm gas within molecular outflows.

\end{itemize}


\acknowledgements I like to thank F. Wyrowski for providing Fig. 1. A
very special thank you to the whole SMA group for making the
instrument possible. I also like to than A. Walsh and Q. Zhang for
comments on the manuscript. H.B. acknowledges financial support by the
Emmy-Noether-Programm of the Deutsche Forschungsgemeinschaft (DFG,
grant BE2578/1).

\end{document}